\documentclass[prb,preprint,superscriptaddress,showpacs]{revtex4}

\usepackage{amsmath,amsfonts,amssymb,bm}
\usepackage{dcolumn}
\usepackage[final]{graphicx}

\newcommand{\modena}[0]{{
  Istituto Nazionale per la Fisica della Materia (INFM) and
  Dipartimento di Fisica,\\
  Universit\`a degli Studi di Modena e Reggio Emilia,
  Via Campi 213/A, 41100 Modena, Italy}}

\newcommand{\graz}[0]{{
  Institut f\"ur Theoretische Physik,
  Karl--Franzens--Universit\"at Graz, Universit\"atsplatz 5,
  8010 Graz, Austria}}

\begin{document}
\bibliographystyle{apsrev}

\title{Electron-hole localization in coupled quantum dots}

\author{Filippo Troiani}
\affiliation{\modena}

\author{Ulrich Hohenester}\email{ulrich.hohenester@uni-graz.at}
\affiliation{\graz}

\author{Elisa Molinari}
\affiliation{\modena}

\date{\today}

\begin{abstract}

We theoretically investigate correlated electron-hole states in vertically coupled quantum dots. Employing a prototypical double-dot confinement and a configuration-interaction description for the electron-hole states, it is shown that the few-particle ground state undergoes transitions between different quantum states as a function of the interdot distance, resulting in unexpected spatial correlations among carriers and in electron-hole localization. Such transitions provide a  direct manifestations of inter- and intradot correlations, which can be directly monitored in experiments.

\end{abstract}

\pacs{73.21.La,71.35.-y,03.67.-a}
\maketitle



Semiconductor quantum dots (QDs) are solid state nanostructures which
allow confinement of carriers in all directions within dimensions
smaller than their de Broglie wavelength.~\cite{hawrylak:98,bimberg:98} 
Quantum confinement results in a
characteristic discrete energy spectrum and a $\delta$-like density of
states; therefore QDs are often referred to as ``artificial atoms''.
Coupling between dots is now becoming an issue of
crucial importance. On the one hand, it is an inherent feature of any
high-density QD ensemble, as, e.g., needed for most optoelectronic
applications.~\cite{bimberg:98} On the other hand, it is essential to
the design of (quantum) information devices, for example QD cellular
automata \cite{lent:99} or QD implementations of quantum computation.~\cite{troiani:00} ``Artificial molecules'' formed by two or more
coupled QDs are extremely interesting also from the fundamental point
of view, since the interdot coupling can be tuned far out of the
regimes accessible in natural molecules, and the relative importance of
single-particle tunneling and Coulomb interactions can be varied in a
controlled way. The interacting states of $N$ electrons in a double dot
were studied theoretically \cite{rontani:98,partoens:00,tejedor:00} and
experimentally by tunneling and capacitance experiments,~\cite{austing:97,oosterkamp:97,schmidt:97,blick:98,brodsky:00,austing:01}
and correlations were found to induce coherence effects and novel
ground-state phases depending on the interdot coupling regime.~\cite{rontani:98,partoens:00,tejedor:00,austing:01}


The effects of correlations on the {\em photoexcited electron and hole
system}\/ in {\em coupled dots}\/ are instead still largely unknown, in
spite of their importance for possible novel applications such as
quantum-information processing devices.~\cite{troiani:00,biolatti:00}
Stacked self-organized dots were demonstrated;~\cite{fafard:00} exciton
splitting in a single artificial molecule was observed in the linear
regime, and explained in terms of single-particle level filling of
delocalized bonding and anti-bonding electron and hole states.~\cite{schedelbeck:97,bayer:01}  When a few photoexcited particles are
present, however, Coulomb interactions between electrons and holes adds to
the homopolar electron-electron (hole-hole) interactions.~\cite{singledot}
The correlated ground and excited states will thus be governed by the competition of
these effects, sofar not included in theoretical descriptions of
photoexcited artificial molecules. However, even for the simplest
symmetric double-dot structures basic questions regarding the nature of
electron-hole states are still open. When more than one electron-hole pair is
photoexcited, what is the most favorable spatial arrangement of the
interacting electrons and holes, i.e., are carriers distributed in both
dots or do they prefer localization in the same dot?
Can the structural parameters be tuned in
order to induce given spatial distributions of electrons and holes in
the artificial molecule, thus allowing for a controlled engineering of
the Hilbert space? 
Can electrons and holes tunnel separately or is the system best described in 
terms of excitonic tunneling?



In this Rapid Communication we analyze ground and excited states of realistic double
quantum dots trough a full configuration-interaction
description of the few-particle electron-hole system. 
It is shown that for a given number of excitons the ground
state configuration undergoes non-trivial quantum transitions as a
function of the interdot distance $d$ resulting in unexpected spatial
correlations among carriers, which can be directly monitored in the
optical spectra.


The initial ingredients of our calculations are the single-particle wavefunctions 
$\phi_\mu^m({\bm r})$ and energies $\epsilon_\mu^m$ for electrons ($m=e$) and holes ($m=h$) 
which are obtained from the solutions of the three-dimensional Schr\"odinger equation 
within the envelope-function and effective-mass approximations.~\cite{hawrylak:98} 
We examine a prototypical confinement for two vertically coupled dots, 
which is double-box-like along $z$ and parabolic in the $(x,y)$-plane;~\cite{dot} 
such parabolic lateral confinement is known to mimic the most important features 
of various kinds of self-assembled dots and to give results in good agreement with experiment.~\cite{hawrylak:98,bayer:00,rinaldi:96} 
The single-particle  wavefunctions factorize into in-plane and $z$-dependent parts, where the in-plane solutions are the well-known Fock-Darwin states.~\cite{hawrylak:98}
Our approach for the description of the interacting electron-hole states is provided by a full configuration interaction (CI) calculation. Within the Hilbert space of all possible single-particle electron-hole excitations $|\ell\rangle$, the many-body energies $E_\lambda$ and states $|\lambda\rangle$ are obtained for a given number of electrons and holes by direct diagonalization of the Hamiltonian matrix 
$\langle\ell|(H_o+H_c)|\ell'\rangle$,~\cite{hohenester:00,rontani:01} with $H_o$ the single-particle Hamiltonian and $H_c$ the Coulomb term:

\begin{equation}
  H_c=\frac 1 2 \sum_{mn}\int d(\bm{rr}')
  \frac{\psi^\dagger_m({\bm r})\psi^\dagger_n({\bm r'})
  \psi_n({\bm r}')\psi_m({\bm r})}{\kappa|{\bm r}-{\bm r'}|},
\end{equation}

\noindent which accounts for all possible electron-electron, hole-hole, and electron-hole interactions; here the field operators $\psi^\dagger_m({\bm r})$ create an electron or hole at position $\bm r$, and $\kappa$ is the semiconductor dielectric constant. In our calculations truncation of the Hilbert space to the 12
single-particle states of lowest energy for both electrons
and holes is found to yield very accurate convergence for the lowest 
few-particle states.



Let us first analyze the single-particle (SP) ingredients of our coupled-QD model.
Figure 1 shows the energies for
the ``bonding'' (solid line) and ``antibonding'' (dashed line) state of
lowest energy for electrons (left axis) and holes (right axis),
respectively; the insets show the wavefunctions along $z$ for two
selected values of the interdot distance $d$. Because of the heavier
hole mass hole tunneling becomes suppressed at smaller interdot
distances $d$ than electron tunneling; this is also reflected in the
faster decrease of the bonding-antibonding splitting with increasing
$d$.


If we would neglect at the lowest order of approximation Coulomb interactions, 
the few-particle electron-hole
states would be simply obtained by filling successive SP
electron and hole states. However, inclusion of Coulomb interactions gives rise 
to a mixing of different SP states, where the gain in
potential energy through Coulomb correlations is achieved at the price
of populating states with higher SP energies. The few-particle
electron-hole states thus result from the detailed interplay of these
trends, where the relative importance of SP energies and the increase 
of the Coulomb ones varies with 
interdot distance $d$. As will be shown in the following, such interplay can
drastically alter the simple-minded SP picture.


We first consider the case of a single electron-hole pair (exciton).
The diamonds in Fig.~2 show the $d$-dependence of the exciton groundstate. 
An estimate of Coulomb
correlations is provided by the usual electron-hole correlation function

\begin{equation}
  g_{eh}({\bm r}_e,{\bm r}_h) =
  \langle \psi^\dagger_e({\bm r}_e)\psi^\dagger_h({\bm r}_h)
  \psi_h({\bm r}_h)\psi_e({\bm r}_e)\rangle,
\end{equation}

\noindent which gives the probability of finding
an electron at position ${\bm r}_e$ when a hole is at position ${\bm
r}_h$. In the insets of Fig.~2 we plot the spatial average of $g_{eh}$ over the
center-of-mass coordinate ${\bm R}=\frac 1 2({\bm r}_e+{\bm r}_h)$, 
$\bar g_{eh}({\bm r})$,
which instead gives the probability of finding the two particles at the 
relative position ${\bm r}={\bm r}_e-{\bm r}_h$.
Owing to
the strong Coulomb interaction the electron and hole stay together and
$\bar g_{eh}$ is localized around ${\bm r}\approx {\bm 0}$ [insets in
Fig.~2, with crosses at ${\bm r}={\bm 0}$].  Thus, {\em even for the
smallest interdot distance $d$ exciton tunneling dominates over
separate electron and hole tunneling}.\/ Note that in a pure
single-particle picture, with electrons and holes occupying the lowest 
available SP states,
there would be no spatial correlation between
them; this would be reflected in $\bar g_{eh}(x,0,z)$ by
a structure with two peaks around $z=\pm d$ and a larger one centred on 
$z=0$ (for an  electron in the left dot the hole is with equal probability in the left 
or right dot, and vice-versa).

The effects of Coulomb-correlations become more striking when
higher-order interactions ---i.e. beyond single excitons---
are taken into account. In contrast to transport experiments,~\cite{austing:97,oosterkamp:97,schmidt:97,blick:98,brodsky:00,austing:01}
which primarily give information about groundstate properties, optical
spectroscopy allows easy access to excited states. For instance, in
non-linear coherent spectroscopy \cite{bonadeo:98} a strong ``pump''
laser prepares the system in an exciton state, and a weak ``probe''
beam measures the optical transitions to biexciton states (i.e., two
Coulomb-correlated electron-hole pairs). The solid lines in Fig.~2 are
associated with 
the transitions from the exciton groundstate to the lowest 
optically active biexciton states
where the two electron-hole pairs have parallel (gray line) or opposite
(black lines) spin orientations; the thickness of the lines is
proportional to the oscillator strengths of the respective transitions.~\cite{hohenester:00}
As to the general trends of the curves, the most striking features 
are the following:
for parallel spins, because of Pauli blocking the
second electron-hole pair is created in an excited state, as
reflected by the increase of the optical transition energy
with increasing $d$
(``antibonding'' behaviour). On the
contrary, for antiparallel spins the overall $d$-dependence of the
lowest biexciton states shows a ``bonding'' behavior, which reflects the 
predominant occupation of the SP ``bonding'' states. 
More in detail, one can see that at a
critical interdot distance $d_c \approx 2$ nm an
anti-crossing occurs between the two solid black lines, where the 
energetically lowest one picks up all the oscillator strength. 
Besides, at the largest interdot distances, $d\cong 3$ nm,
two of the lines merge with the exciton energy: the corresponding
biexciton states consist of two spatially separated excitons
localized in the two dots, as discussed below.
At the same values of $d$, in the biexciton groundstate 
all four particles are localized in the same dot, where because of Coulomb correlations the energy is reduced by $\approx$1 meV as compared to the uncorrelated case (biexciton binding energy).~\cite{bayer:00,hohenester:00}

In the following we analyze in more detail the lowest biexcitonic
states. As will be apparent from the results of Fig.~3, there exist two major trends for
correlation effects:
In case of ``anti-correlation'' electrons and holes tend to avoid each other, 
and the repulsion  energy is minimized when each dot is populated by one electron-hole pair. 
As in the absence of external fields there is no net charge distribution 
in each QD, there exists an only very weak long-range interaction between the 
excitons---even in regimes 
where the QDs are tunneling-coupled (Fig. 1);
as consequence, the biexcitonic energy at large $d$ 
tends to twice the excitonic one (grey and upper black curve, respectively, in Fig. 2). 
Alternatively, correlation can favour spatial arrangements where all four particles 
are localized in the same QD.
This requires a correlation of higher order \cite{hoc} with occupation of 
the antibonding states for both electrons 
and holes. When all particles are localized in the same QD, 
in-plane spatial correlations becomes effective which give rise to 
a short-range interaction (analogue to a classical interaction between 
induced dipoles) and to the well-known
biexcitonic binding energy $\Delta E_{2X}$. 
This spatial arrangement becomes energetically more 
favourable with respect to the former when the cost of its 
stronger degree of correlation in terms of SP energy 
is $\lesssim \Delta E_{2X}$.

Figure 3 shows the
$d$-dependence of the (a) energies, and the mean (b) electron-hole, (c)
electron-electron, and (d) hole-hole distances for the four 
biexciton states of lowest energy. 
Let us first focus on the ground state.
At the smallest values of $d$ , the 
ground state is seen to correpond to: no electron-hole correlation, 
intermediate anti-correlation between electrons and a pronounced one between holes
[left insets in Figs. 3 (b-d)]. This different degree of correlation
for electrons and holes 
can be attributed to the different SP level splittings (Fig.~1) and the 
corresponding different cost in SP energy to create two-particle correlations. 
Apparently, at $d_c$
the system undergoes an abrupt transition. The correlation functions 
[middle and right insets of Figs. 3 (b-d)] show how spatial correlation becomes
more and more effective, with strong 
correlations between all particles. Such transition is also confirmed 
by the trends of the mean carrier-carrier distances that abruptly 
decrease for $d \approx d_c$ . The second 
optically active biexcitonic state (continuous grey lines) 
shows a precisely reversed trend (and a similar anti-crossing occurs between the 
two intermediate,
non optically active states [dotted lines in Figs. 3(a-d)].

Further calculations show that the distance at which the transition
occurs can be tailored by controlling the QD confinement and interdot
coupling. In addition, we have found that the effects are preserved in
the presence of small asymmetries between both dots, although the
transitions occur at different distance $d$ for different states, and
that similar phenomena persist for a larger number of photoexcited
particles in the double dot.~\cite{troiani:01}.


In conclusion, we have addressed for the first time few-particle
correlations in photoexcited coupled quantum dots. We have demonstrated
that they can induce novel transitions between different quantum states
as a function of the interdot distance $d$, and are essential for a
quantitative or even qualitative description of the electron and hole
states. The fundamental reason for these transitions lies in a very general 
phenomenon, namely that interdot coupling controls the balance between 
the gain in Coulomb energy obtained by the spatial correlation 
of carriers (which requires occupation of higher-lying single-particle 
states) and the corresponding cost in single-particle energy.
These transitions leave a clear fingerprint in the nonlinear optical
spectra. Beside their
fundamental interest as manifestations of inter- and intradot
correlations, these quantum transitions may be used to tailor the
Hilbert space structure and the qubit identification in quantum
computation schemes based on coherent optical control of coupled
quantum dots.

This work has been supported in part by INFM through PRA-99-SSQI, and
by the EU under the TMR Network ``Ultrafast Quantum Optoelectronics''
and the IST programme ``SQID''.


\begin{figure}
\centerline{\includegraphics[width=0.65\columnwidth]{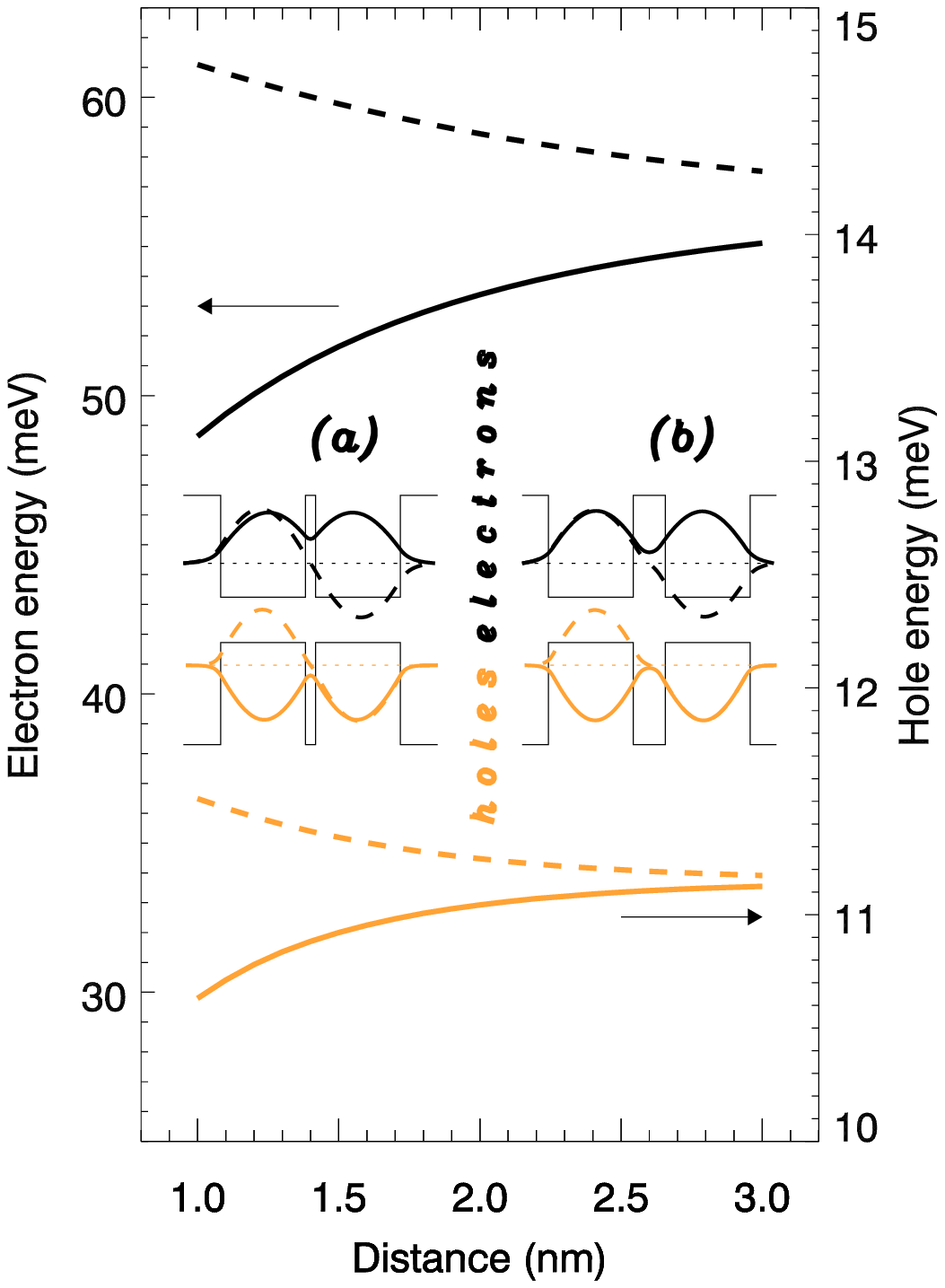}}
\caption{
Single-particle energies for electrons (left axis) and holes (right
axis) in a symmetric double QD   
as a function of the interdot distance $d$. The curves correspond
to the lowest Fock-Darwin state and to the bonding (solid line) and
antibonding (dashed line) groundstates along $z$. The insets show the
electron and hole wavefunctions along $z$ for two selected interdot
distances of 1.2 nm [(a)] and 2.8 nm [(b)], respectively.
}
\end{figure}

\begin{figure}
\centerline{\includegraphics[width=0.65\columnwidth]{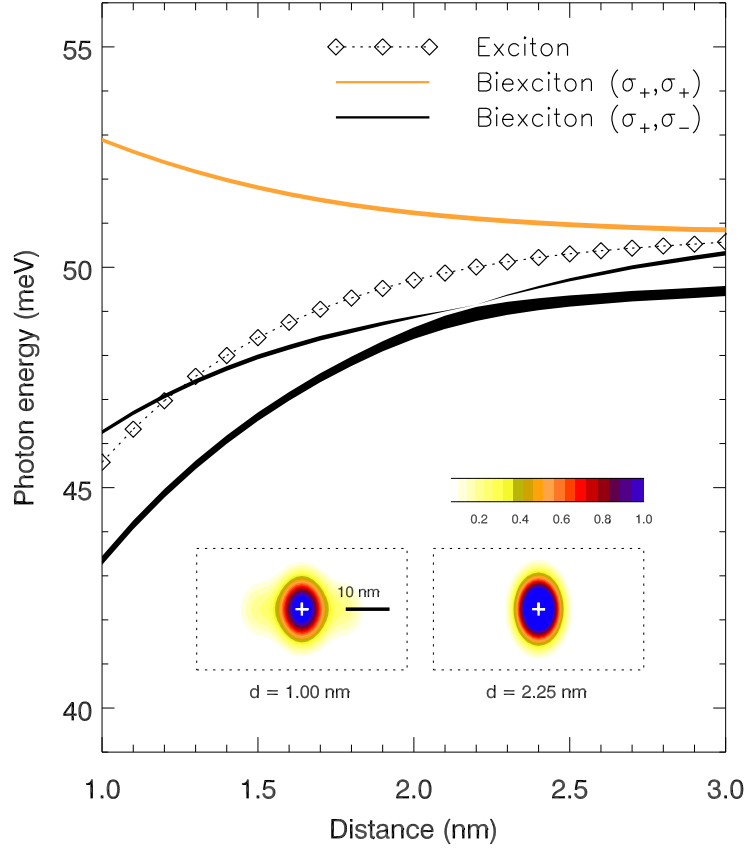}}
\caption{
Excitation energies of a symmetric double QD.  
Diamonds: Exciton energies as a function of interdot distance $d$. Solid
lines: Transition energies from the exciton groundstate to the
biexciton state where the two electron-hole pairs have parallel (gray
line) or antiparallel spins (black lines); the thickness of the lines
corresponds to the oscillator strengths of the corresponding
transitions. Insets: Normalized electron-hole correlation function $\bar
g_{eh}(x,0,z)$ ($x$ and $z$ in vertical and horizontal directions,
respectively) for the exciton groundstate; the crosses indicate
$x=z=0$.
}
\end{figure}

\begin{figure}
\centerline{\includegraphics[width=0.45\columnwidth]{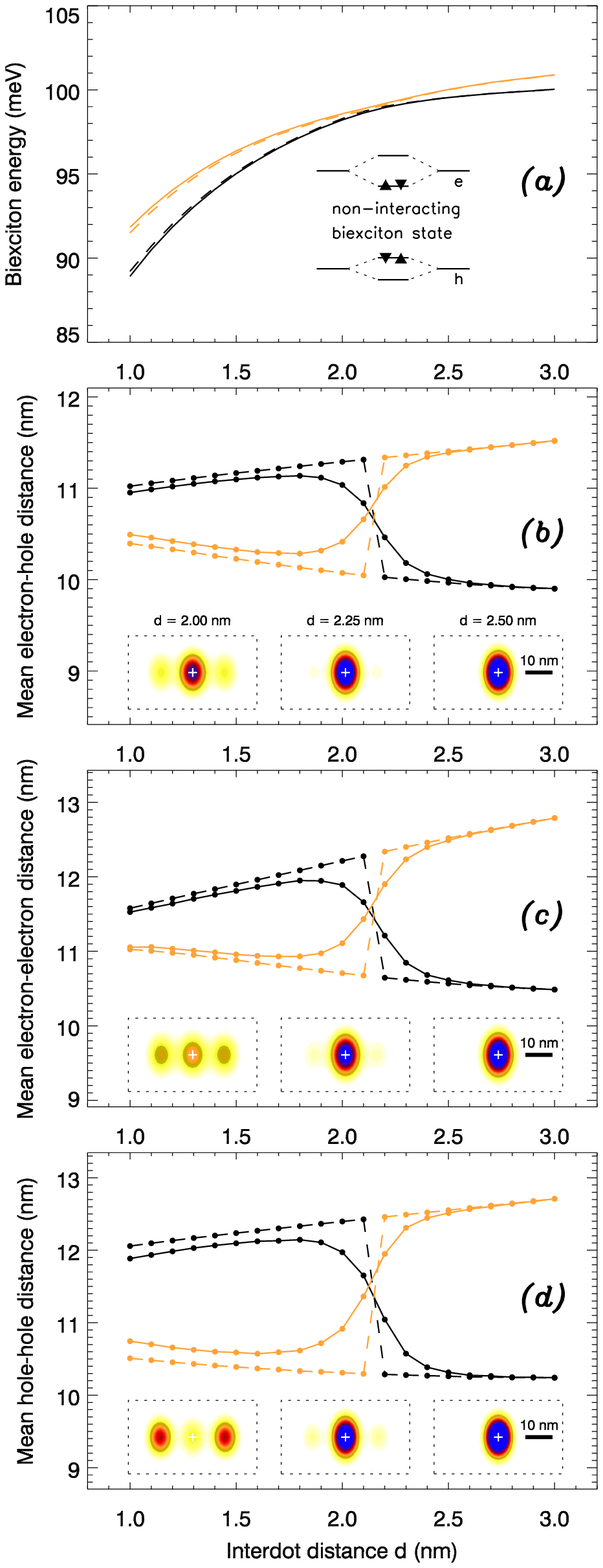}}
\caption{
Biexciton energies [(a)] and mean distances between particles
[(b--d)] in a symmetric double QD. The four lowest energetic levels 
are considered. The continuos curves refer to 
the optically active biexciton states involved in the transitions 
of Fig.~2 (black lines).
Insets: Electron-hole [(b)], 
electron-electron [(c)] and hole-hole [(d)] correlation functions $\bar
g(x,0,z)$. 
}
\end{figure}


\begin{thebibliography}{10}

\bibitem{hawrylak:98}
L. Jacak, P. Hawrylak, and A. Wojs, 
{\em Quantum Dots}\/ (Springer, Berlin, 1998).

\bibitem{bimberg:98}
D. Bimberg, M. Grundmann, and N. Ledentsov, 
{\em Quantum Dot Heterostructures}\/ (John Wiley, New York, 1998).

\bibitem{lent:99} G.L. Snider {\em et al.},\/ J. Appl. Phys. {\bf 85}, 
4283 (1999).

\bibitem{troiani:00} F. Troiani, U. Hohenester, E. Molinari, Phys. Rev. B
{\bf 62}, R2263 (2000), and references therein.

\bibitem{rontani:98}
M. Rontani, F. Rossi, F. Manghi, E. Molinari, 
Solid State Comm. {\bf 112}, 151 (1998).

\bibitem{partoens:00}
B. Partoens and F.M. Peeters, Phys. Rev. Lett. {\bf 84}, 4433 (2000).

\bibitem{tejedor:00}
L. Martin-Moreno, L. Brey, and C. Tejedor, 
Phys. Rev. B{\bf 62}, R10633 (2000).

\bibitem{austing:97} D.G. Austing et al., Jap.
J. Appl. Phys. {\bf 36}, 1667 (1997); Semicond. Sci. Technol. {\bf 12},
631 (1997); D.G. Austing et al., Physica B {\bf 251}, 206 (1998).

\bibitem{oosterkamp:97}
Phys. Rev. Lett. {\bf 80}, 4951 (1998); Nature  {\bf 395}, 873 (1998); T. Fujisawa et 
al., Science {\bf 282}, 932 (1998).

\bibitem{schmidt:97}
T. Schmidt, R. J. Haug, K. v. Klitzing, A. F\"orster, and H. L\"uth,
Phys. Rev. Lett. {\bf 78}, 1544 (1997).

\bibitem{blick:98}
R. H. Blick {\em et al.}, Phys. Rev. Lett. {\bf 80}, 4032 (1998); 
{\em ibid.} {\bf 81}, 689 (1998).

\bibitem{brodsky:00}
M. Brodsky, N.B. Zhitenev, R.C. Ashoori, L.N. Pfeiffer, and K.W. West,
Phys. Rev. Lett. {\bf 85}, 2356 (2000).

\bibitem{austing:01} 
M. Pi {\em et al.}, Phys. Rev. Lett. {\bf 87}, 66801 (2001).

\bibitem{biolatti:00}
E. Biolatti, R. C. Iotti, P. Zanardi, and F. Rossi,
\prl {\bf  85}, 5647 (2000).

\bibitem{fafard:00}
S. Fafard, M. Spanner, J. P. McCaffrey, and Z. R. Wasilewski, 
Appl. Phys. Lett. {\bf 75}, 2268 (2000), and references therein.

\bibitem{schedelbeck:97}
G. Schedelbeck, W. Wegschreider, M. Bichler, G. Abstreiter,
Science {\bf 278}, 1792 (1997).

\bibitem{bayer:01}
M. Bayer et al., Science {\bf 291}, 451 (2001).

\bibitem{singledot} For single quantum dots it is now known from
previous theoretical and experimental work that few-particle Coulomb
correlations dominate the optical spectra in the non linear regime.
See, e.g., A. Hartmann, Y. Ducommun, E. Kapon, U. Hohenester, and E.
Molinari, Phys. Rev. Lett. {\bf 84}, 5648 (2000), and references
therein.

\bibitem{dot} In our calculations we use a symmetric double-box-like 
structure along $z$ and a parabolic confinement in lateral directions.
The dot and material parameters are: 10 nm for the box widths, 
400 meV and 215 meV for the depth of the wells for electrons and holes, respectively; 
the effective masses are $m^{(e)}=0.067 m_{0}$ and 
$m^{(h)}=0.38 m_{0}$ (material parameters for GaAs/AlGaAs).
As to the in-plane parabolic potential, $\omega_o^{(e)}=20$ meV for 
electrons, and $\omega_o^{(h)}=3.5$ meV for holes. With such values of
the parameters, electron and hole wavefunctions have the same 
lateral extension.  

\bibitem{bayer:00}
M. Bayer, O. Stern, P. Hawrylak, S. Fafard, and  A. Forchel,
Nature {\bf 405}, 923 (2000).

\bibitem{rinaldi:96}
R. Rinaldi {\em et al}.\/,
Phys. Rev. Lett. {\bf 77}, 342 (1996);
Phys. Rev. B {\bf 57}, 9763 (1998); 
Phys. Rev. B {\bf 62}, 1592 (2000).

\bibitem{hohenester:00}
U. Hohenester and E. Molinari,
phys. stat. sol. (b) {\bf 221}, 19 (2000).

\bibitem{rontani:01}
M. Rontani, F. Troiani, U. Hohenester, and E. Molinari,
Solid State Commun. {\bf 119}, 309 (2001).

\bibitem{hoc}
Here the overall wavefunction can no longer be written approximately as 
the electron wavefunction times the hole wavefunction.


\bibitem{bonadeo:98}
N. H. Bonadeo {\em et al.},\/ Phys. Rev. Lett. {\bf 81}, 2759 (1998).

\bibitem{troiani:01}
F. Troiani, U. Hohenester and E. Molinari, to be published. 

\end{thebibliography}
\end{document}